\documentclass[twocolumn,a4paper,10pt]{article}
\usepackage{graphicx}
\begin{document}
\twocolumn[
\begin{center}
\Huge{Equations of Motion of Systems with Internal Angular Momentum}
\end{center}
\small
%
%
\renewcommand{\thefootnote}{\fnsymbol{footnote}}
{\bf Manuel Dorado}{\footnote[1]{E-mail:  mdorado@cayacea.com}}  \\
{\emph CENITA, S.A., Miguel Yuste, 12, 28037 Madrid, Spain}   \\
{\bf Abstract} \\
Using the Euler's equations and the Hamiltonian formulation, an attempt
has been made to obtain the equations of motion of systems with
internal angular momentum that are moving with respect to a reference
frame when subjected to an interaction. This interaction involves the
application of a torque that is permanently perpendicular to the
internal angular momentum vector. \\
]

{
\hspace*{1.cm}

\section*{\normalsize INTRODUCTION}

\renewcommand{\thefootnote}{}
\footnote[1]{$^*$E-mail:  mdorado@cayacea.com}
\renewcommand{\thefootnote}{\arabic{footnote}}
The angular momentum of a many-particle system with respect to its centre 
of mass is known as the "internal angular momentum" and is a property of
the system that is independent of the observer. Internal angular momentum is 
therefore and attribute that characterizes a system in the same way as its
mass or charge. In the case of a rigid body and particularly in the case of
an elementary particle, the internal angular momentum is also referred to
as "spin".

To the author's knowledge, the dynamical behaviour of systems with internal
angular momentum has not been systematically studied within the framework of
classical mechanics (See References [1]). This paper approaches this type of
problem through a model comprising a rotating cylinder, with constant velocity
of rotation, around its longitudinal axis. The equations of motion are 
obtained using the Euler's equations and by the Hamiltonian procedure.

Results coincide when the problem is solved using vectorial algebra or 
Lagrangian formalism (unpublished). However, as a result of changes in 
"perspective", each method uncovers new peculiarities regarding the intimate
nature of the system's behaviour.

\section*{\normalsize FORMULATION OF THE PROBLEM}

Let us consider a rigid cylindrical solid with internal angular momentum
(with respect to its centre of mass) $\vec{L}$, whose centre of mass moves
at a constant velocity $\vec{\nu}$ with respect to a reference frame which can
be defined as soon as the interaction in the cylinder starts. It is aimed
to obtain the equations of motion to describe the dynamical behaviour of
the system from the instant that it undergoes an interaction, by applying
a torque $\vec{M}$ that is permanently perpendicular to the internal angular
momentum (See Fig.1).

When solving the problem, the following points are taken into account:
\begin{enumerate}
\item[(a)] The cylinder will continue spinning about its longitudinal axis
at a constant angular velocity $\vec{\omega}$ throughout all the movement.
In other words, its internal angular momentum module is constant. The energy
that the cylinder possesses as a result of its rotation about its longitudinal
axis is consequently considered to be internal and does not interfere with
its dynamical behaviour.
\item[(b)] The derivative of the internal angular momentum, $\vec{L}$, with
respect to a frame of axes of inertial reference ($X$, $Y$, $Z$) satisfies the
equation
\begin{equation}
\left(\frac{d\vec{L}}{dt}\right)_{XYZ} = \left(\frac{d\vec{L}}{dt}\right)_{X'Y'Z'} + \vec{\Omega} \times \vec{L}
\end{equation}
in which $\vec{\Omega}$ is the rotation velocity of the frame linked to the
solid ($X'$, $Y'$, $Z'$) about the frame of inertial reference axes ($X$, $Y$, $Z$).

\item[(c)] Only infinitesimal motions are considered that are compatible with
the system's configurational variations brought about by the applied torque.

\item[(d)] The total energy of the system is not explicitly dependent on time.

\item[(e)] No term is included that refers to the potential energy. According
to hypothesis, the applied torque is a null force acting on the system and the
possibility of including a potential from which the applied torque derives is
unknown.

\end{enumerate}

\includegraphics[width=7cm, height=5cm]{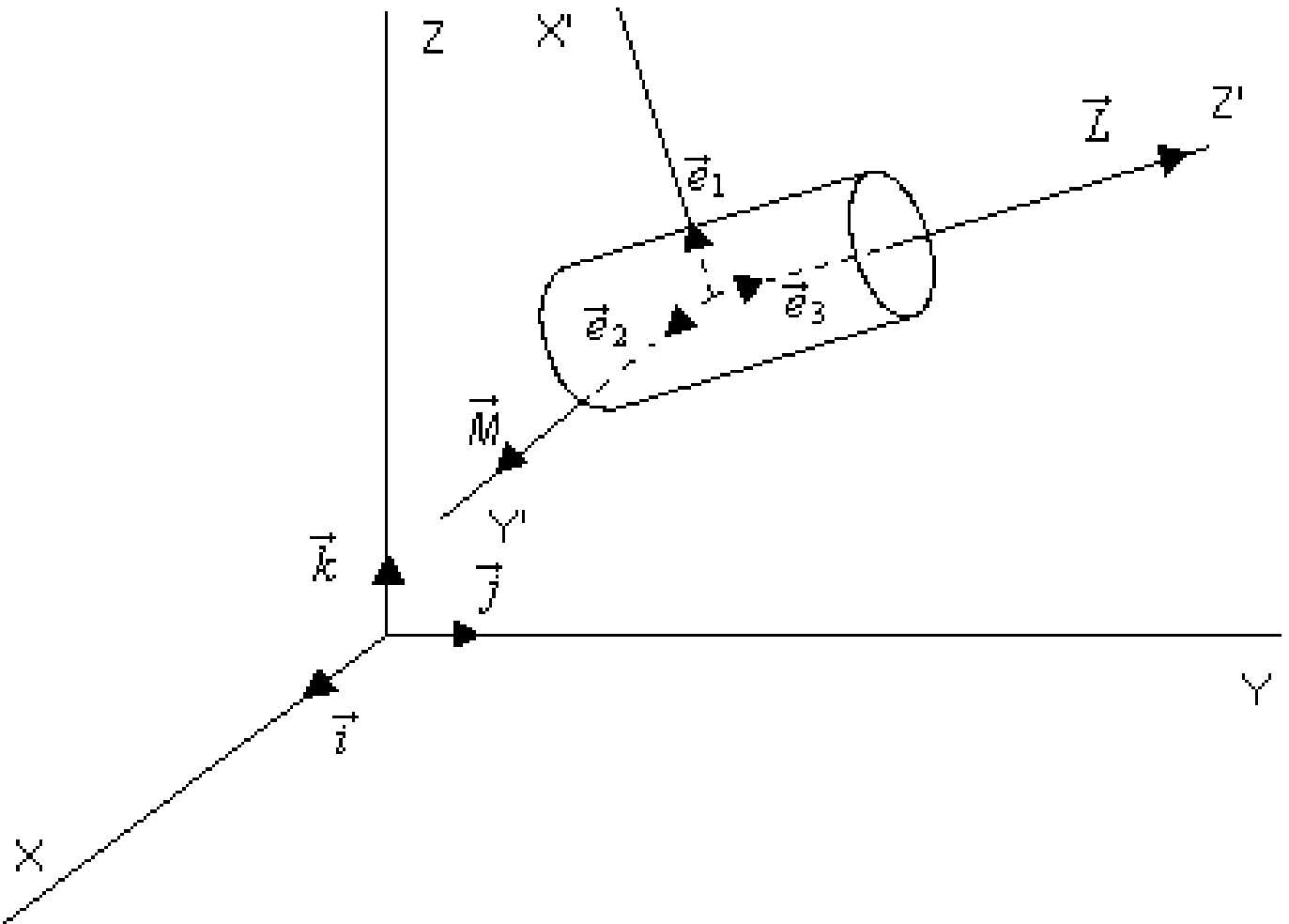}
\begin{center}
\footnotesize{Figure 1. Formulation of the problem}
\end{center}
\section*{\normalsize HOW TO APPROACH THESE PROBLEMS THROUGH THE EULER'S EQUATIONS?}

The figure 1 shows a cylindrical rigid body with angular momentum, $\vec{L}$, 
about its longest axis. We should consider two referential frames: one associated
with the solid ($X'$, $Y'$, $Z'$), having the $Z'$ axis on the direction of the body's
longest axis. We will refer this system of coordinates to a inertial frame
($X$, $Y$, $Z$).

In a given instant, a torque $\vec{M}$ is applied on the rigid body.

We call $\vec{\Omega}$ to the rotational velocity of the system of coordinates
associated with the solid ($X'$, $Y'$, $Z'$), viewed from the inertial frame ($X$, $Y$, $Z$).
From here, we will consider the system of coordinates ($X'$, $Y'$, $Z'$) all along
the work.

Assuming this situation, it is known that
\begin{equation}
\left(\frac{d\vec{L}}{dt}\right)_{XYZ} = \left(\frac{d\vec{L}}{dt}\right)_{X'Y'Z'} + \vec{\Omega} \times \vec{L}
\end{equation}

It is clear that we have two different velocities acting:

\begin{enumerate}
\item[(a)] $\vec{\omega}$: the rotational velocity of the body around its
longest axis (let us call it the intrinsic rotational velocity of the body).

\item[(b)] $\vec{\Omega}$: the rotational velocity of the system of coordinates
associated with the body ($X'$, $Y'$, $Z'$) viewed from the frame of inertia ($X$, $Y$, $Z$).
\end{enumerate}

We can write $\vec{L}$ and $\vec{M}$, refered to ($X'$, $Y'$, $Z'$), as follows:

\begin{eqnarray}
\vec{L} & = & I_1 \omega_1 \vec{e}_1 + I_2 \omega_2 \vec{e}_2 + I_3 \omega_3 \vec{e}_3 \\
\vec{\Omega} & = & \Omega_1 \vec{e}_1 + \Omega_2 \vec{e}_2 + \Omega_3 \vec{e}_3 \\
\vec{M} & = & M_1 \vec{e}_1 + M_2 \vec{e}_2 + M_3 \vec{e}_3
\end{eqnarray}

So, the equation (2) takes the expression
\begin{eqnarray}
I_1 \dot{\omega}_1 + I_3 \omega_3 \Omega_2 - I_2 \omega_2 \Omega_3 & = & M_1 \nonumber \\
I_2 \dot{\omega}_2 + I_1 \omega_1 \Omega_3 - I_3 \omega_3 \Omega_1 & = & M_2 \\
I_3 \dot{\omega}_3 + I_2 \omega_2 \Omega_1 - I_1 \omega_1 \Omega_2 & = & M_3 \nonumber
\end{eqnarray}

These equations are known as \emph{modified Euler's equations} and can also be
found in (1.e).

In the case treated above, we must substitute the angular momentum
\begin{equation}
\vec{L} = I_3 \omega_3 \vec{e}_3
\end{equation}
in (6):
\begin{eqnarray}
I_3 \omega_3 \Omega_2 & = & M_1 \nonumber \\
I_3 \omega_3 \Omega_1 & = & M_2 \\
I_3 \dot{\omega}_3 & = & M_3 \nonumber
\end{eqnarray}

And then we get
\begin{eqnarray}
\Omega_2 & = & \frac{M_1}{I_3 \omega_3} \nonumber \\
\Omega_1 & = & \frac{M_2}{I_3 \omega_3} \\
\dot{\omega}_3 & = & \frac{M_3}{I_3} \nonumber
\end{eqnarray}

If the torque is on the axis $Y'$, $M_1$ = 0, $M_3$ = 0 and:
\begin{eqnarray}
\Omega_2 & = & 0 \nonumber \\
\Omega_1 & = & \frac{M_2}{I_3\omega_3} \\
\dot{\omega}_3 & = & 0 \nonumber
\end{eqnarray}

Where $\Omega_1$ coincides with the known velocity of precession.

We have obtained that the body moves along a circular trajectory with a
rotational velocity $\Omega_1$, but this is not enough to determinate the
radius of the orbit.

The principle of conservation for the energy should lead us to the expression
for this radius.

Assuming our initial hypothesis, the forces are applied perpendicular to the
plane of movement of the solid and so, these forces do not produce any work
while the cylinder is moving.

We have two conditions that have to be satisfied by the body while rotating:
\begin{enumerate}
\item[(a)] $\Omega = \frac{M}{L}$

\item[(b)] The total energy should remain constant.
\end{enumerate}

Before the torque is applied the energy of the cylinder takes the following
expression:
\begin{equation}
E = \frac{1}{2} m \nu_o^2 + \frac{1}{2}I \omega_0^2
\end{equation}

When the torque is acting on the cylinder about the line defined by the unit
vector $\vec{e}_2$, it rotates along a trajectory, generally defined by $r(t)$
and $\dot{\theta}(t)$ that, in this case, coincides with $\Omega(t)$ and its
energy takes the following expression:
\begin{equation}
E = \frac{1}{2}m\left(\dot{r}^2 + r^2 \dot{\theta}^2\right) + \frac{1}{2} I \omega_0^2
\end{equation}

In this particular case our axes are principal axes of inertia, therefore we have
$\frac{\delta L_{X^{\prime}}}{\delta t} = M_1$, and then,
$\frac{\delta}{\delta t}(m r^2 \dot{\theta}) = 0$.

And therefore $m r^2 \dot{\theta} = const$.

In the particular case in which $\dot{\theta} = \Omega$ is a constant, $r$ has
to be a constant (that is, $\dot{r}$ has to be zero) in order to satisfy this
equation. We can conclude that the body moves in a circular orbit. The energy
takes the following expression:
\begin{equation}
E = \frac{1}{2} m r^2 \Omega^2 + \frac{1}{2}I \omega_0^2
\end{equation}

Both expressions (11) and (13) have to represent the same energy. Therefore,
comparing the two expressions we can obtain the radius of the orbit:
\begin{equation}
r = \frac{\nu_0}{\Omega}
\end{equation}

Looking at the results arising from the previous study, we can say that the
body rotates following a circular orbit, with a radius given by the last
expression. This circular trajectory implies that the velocity vector must
precess jointly with the intrinsic angular momentum vector. From this last
conclusion it can be proved that the force needed to cause the particle to
draw a circular trajectory is expressed as:
\begin{equation}
\vec{F} = m \vec{\nu} \times \vec{\Omega}
\end{equation}

When the internal angular moment of the system does not coincide with one of
the principal axes of inertia, the treatment of the problem is much more
complex, but as it will be seen in the next section, the results given above
area completely general.

\section*{\normalsize HAMILTONIAN FORMULATION. EQUATIONS OF MOTION AND THEIR SOLUTION}

The equations of motion are obtained by the Hamiltonian formulation, where
the independent variables are the generalized coordinates and moments. To do this,
a basis change of the frame ($q$, $\dot{q}$, $t$) to ($p$, $q$, $t$) is made
using the Legendre transformation.

From the function
\begin{equation}
H(p, q, t) = \sum_i \dot{q}_i p_i - L(q, \dot{q}, t)
\end{equation}
a system of 2$n$ + 1 equations is obtained.
\begin{eqnarray}
\dot{q}_i= \frac{\partial H}{\partial t}; & & -\dot{p}_i = \frac{\partial H}{\partial q_i}; \nonumber \\
\\
-\frac{\partial L}{\partial t} & = & \frac{\partial H}{\partial t} \nonumber
\end{eqnarray}

If the last equation in (17) is excluded, a system of 2$n$ first order
equations is obtained, known as canonical Hamilton equations.

By substituting $H$, the first order equations of motion, are obtained.

It should be noted that the Hamiltonian formulation is developed for holonomous
systems and forces derived from a potential that depends on the position or
from generalized potentials. A torque is applied to the present system. The
result of the forces is zero on the centre of mass and, therefore, it is meaningless
to refer to potential energy.

On the other hand, the Hamiltonian function concept does remain meaningful.

By using polar coordinate in the movement plane,
\begin{equation}
\nu_r = \dot{r}; \hspace{1cm} v_{\theta} = r \dot{\theta}
\end{equation}

The Lagrangian is
\begin{equation}
L = \frac{1}{2} m (\dot{r}^2 + r^2 \dot{\theta}^2 )
\end{equation}

The generalized moments are
\begin{equation}
P_r = m \dot{r}; \hspace{1cm} P_{\theta} = m r^2 \dot{\theta}
\end{equation}
so that
\begin{equation}
\dot{r} = \frac{P_r}{m}; \hspace{1cm} \dot{\theta} = \frac{P_{\theta}}{m r^2}
\end{equation}    

The Hamiltonian $H$ is introduced using the equation
\begin{equation}
H(p, q, t) = \sum_i \dot{q}_i p_i - L(q, \dot{q}, t)
\end{equation}
resulting in
\begin{displaymath}
H = P_r \dot{r} + P_{\theta} \dot{\theta} - \left(\frac{1}{2} m \dot{r}^2 + \frac{1}{2} m r^2 \dot{\theta}^2 \right)
\end{displaymath}

If the generalized velocities are substituted by the generalized moments, the
following is obtained
\begin{equation}
H = \frac{P_r^2}{2m} + \frac{P^2_{\theta}}{2 m r^2}
\end{equation}

The first pair of Hamilton equations is
\begin{displaymath}
\dot{r} = \frac{\partial H}{\partial P_r} = \frac{P_r}{m}; \hspace{1cm} \dot{\theta} = \frac{\partial H}{\partial P_{\theta}} = \frac{P_{\theta}}{m r^2}
\end{displaymath}

The second pair of equations is
\begin{displaymath}
-\dot{P}_r = \frac{\partial H}{\partial r} = -\frac{P^2_{\theta}}{m r^3}; \hspace{1cm} -\dot{P}_{\theta} = \frac{\partial H}{\partial \theta} = 0
\end{displaymath}                                                                                                        

The second of these equations shows that the angular momentum $\mathit{J}$ is
conserved.
\begin{equation}
P_{\theta} = \mathit{J} = const.
\end{equation}

The first gives the radical equation of motion,
\begin{equation}
\dot{P}_r = m \ddot{r} = \frac{J^2}{m r^3}
\end{equation}
as $J = P_{\theta} = m r^2\dot{\theta}$, by substituting it results that
\begin{equation}
\dot{P}_r =  \frac{m^2 r^4 \dot{\theta}^2}{m r^3} = m r \dot{\theta}^2
\end{equation}

The term $-\frac{\partial V}{\partial r}$ normally appears in this radial
equation of motion and represents the force derived from a potential. In
other words,
\begin{equation}
m \ddot{r} = m r \dot{\theta}^2 - \frac{\partial V}{\partial r}
\end{equation}

In the present case, this term does not exist.

To complete one's knowledge of the system's development, those relations
derived from the constraints, have to be resorted to.
\begin{enumerate}
\item $E = \frac{1}{2}m\nu^2 = const. \Rightarrow |\nu| = const$

\item $\left(\frac{d \vec{L}}{dt}\right)_{XYZ} = \left(\frac{d \vec{L}}{dt}\right)_{X'Y'Z'} + \vec{\Omega} \times \vec{L}$
\end{enumerate}
where the applied external torque
\begin{equation}
\vec{M} = \left(\frac{d \vec{L}}{dt}\right)_{XYZ}
\end{equation}
by hypothesis it is known that \\
$\left(\frac{d \vec{L}}{dt}\right)_{X'Y'Z'} = \vec{0}$ and from (1) it is concluded that   \\
$\left(\frac{d \vec{L}}{dt}\right)_{X'Y'Z'} = \vec{\Omega} \times \vec{L}$ \\
and it is obtained that $\Omega = \frac{M}{L}$.

This vector represents the rotation velocity of the frame of axes linked to the
cylinder ($X'$, $Y'$, $Z'$) about the frame of inertial references axes
($X$, $Y$, $Z$).

In polar coordinates, the variable defining the rotation about the frame of
axes is $\dot{\theta}$ and it is concluded that
\begin{equation}
\dot{\theta} = \Omega = \frac{M}{L}
\end{equation}
substituting in the radial equation of motion, results that
\begin{equation}
m \ddot{r} = m r \dot{\theta}^2 = m r \Omega^2
\end{equation}

Moreover, $P_{\theta}$=const. and at the initial moment equals
\begin{equation}
P_{\theta} = m r^2 \dot{\theta} = m r \nu
\end{equation}
as $\nu$ is constant, it is concluded that $r$ is also constant. Finally it is
found that \\
\begin{equation}
r = \frac{\nu}{\Omega} \qquad \mbox{and} \qquad  m \ddot{r} = m \nu \Omega
\end{equation}

\subsection*{Further considerations}

The Hamiltonian is independent of $\theta$. This is an expression of the
system's rotation symmetry or, in other words, there is no preferred alignment
in the plane.

The equation $\frac{\partial H}{\partial \theta} = 0$ means that the energy of
the system remains invariable if it is turned to a new position without changing
$r$, $P_r$ or $P_{\theta}$. This aspect is clearly proved.

Worthy of mention in this case is the fact that the Hamiltonian equations do not
provide, in this type of problem, complete information on the radial movement.
The reason for this is that the central force is a function of $\theta$ which
is a coordinate that does not appear in the Hamiltonian (it can be ignored) and
can only be calculated using the constraints.

This proposal is valid for any property of the particle that displays vectorial
characteristics and is sensitive to an external torque of the type described,
without the need to ascertain the real physical nature of that property. The 
proposal can be extended to equivalent problems, despite their not involving the
presence of internal angular momentum as a central characteristic.

Poisson brackets. The Poisson bracket $[P_{\theta}, H]$ is
\begin{displaymath}
[P_{\theta}, H] = \left[\frac{\partial P_{\theta}}{\partial \theta}\frac{\partial H}{\partial P_{\theta}} - \frac{\partial P_{\theta}}{\partial P_{\theta}}\frac{\partial H}{\partial \theta}\right] = - \frac{\partial H}{\partial \theta}
\end{displaymath}
as $\frac{\partial H}{\partial \theta} = 0$, it is concluded that $[P_{\theta}, H] = 0$.

\section*{\normalsize CHARGED PARTICLE IN A MAGNETIC FIELD}

Let us apply this theory to the specific case of a charged particle with spin
$\vec{s}$ and magnetic momentum $\vec{\mu}$, moving at a velocity $\vec{\nu}_0$ in
a uniform magnetic field of flux density $\vec{B}$.

As it's well known, the interaction between $\vec{B}$ and $\vec{\mu}$ makes a
torque to act upon the particle. This situation verifies every condition we
have defined in our hypothesis, and so, the behaviour of the particle must be
in agreement with the theoretical model we have developed above.

Considering the most general of the cases, let us suppose that $\vec{\mu}$ forms
unknown angles $\theta$ with $\vec{B}$ and $\alpha$ with $\vec{\nu}_0$. This
situation is represented in the following figure (See Fig. 2).

\includegraphics[width=7cm, height=5cm]{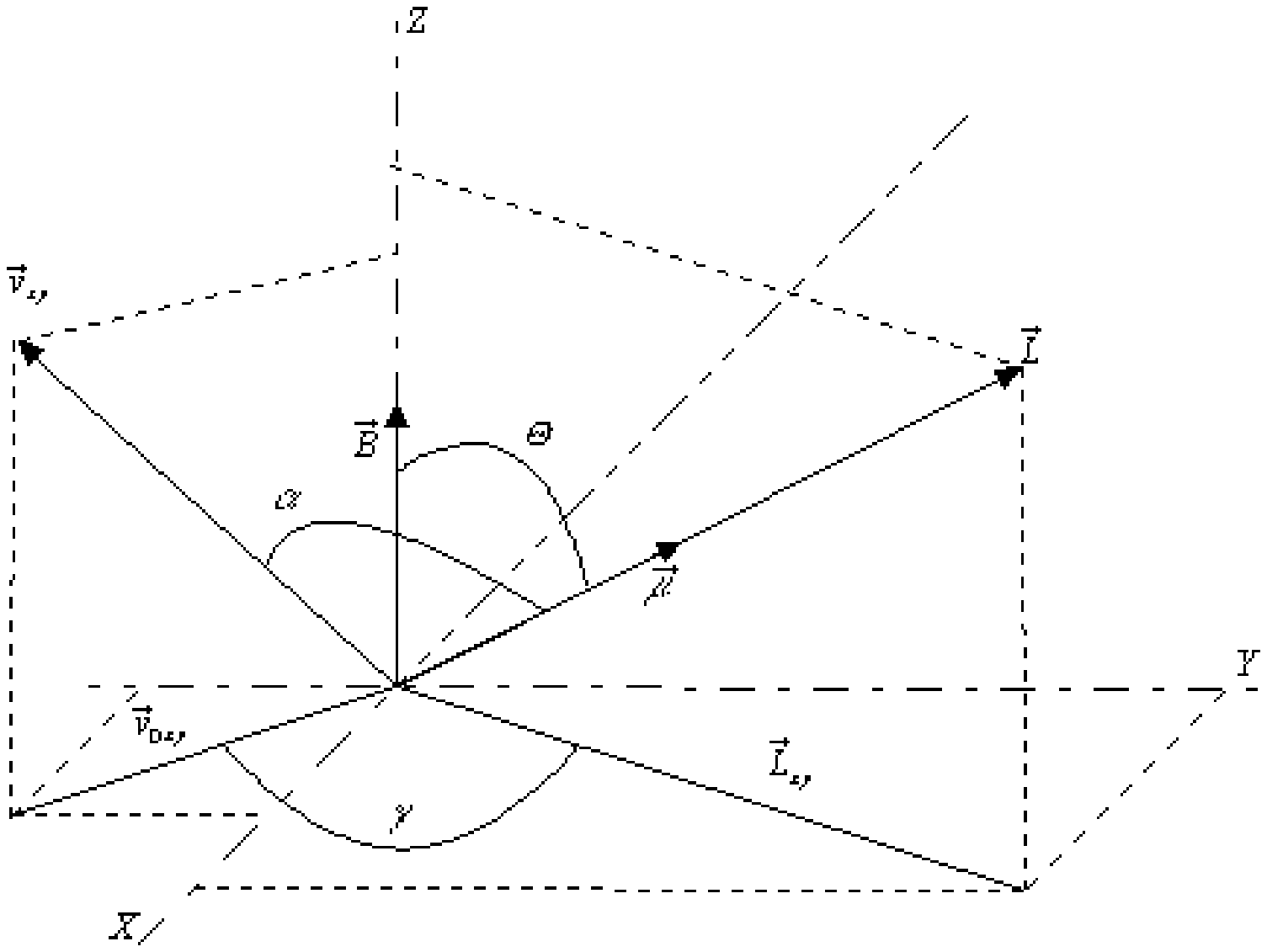}
\begin{center}
\footnotesize{Figure 2. Charged particle in a magnetic field.}
\end{center}

It is clear that the rotation of $\vec{\nu}_0$ causes rotation of the
$\vec{\nu}_0$ component on the $XY$ plane, $\nu_{0xy}$. Furthermore,
according to the theory here presented, the particle will trace a helix
and the $\vec{\nu}_0$ component in the direction of the $Z$ axis, $\vec{\nu}_{0z}$,
will not be affected.

In this case the angular momentum $\vec{L} = \vec{s}$.

If the particle stops inside $\vec{B}$, the rotational velocity of the angular
momentum would be expressed by:
\begin{equation}
\Omega = \frac{|\vec{\mu} \times \vec{B}|}{|\vec{L}|} = \frac{|\vec{M}|}{|\vec{L}|}
\end{equation}

Obviously, if the angle $\alpha$ formed by $\vec{s}$ and $\vec{\nu}_0$ remains
constant, the angle $\gamma$ formed by $\vec{s}_{xy}$ and $\vec{\nu}_{0xy}$ will
also remain constant. The velocity of $\vec{\nu}_{axy}$ can be calculated merely
by calculating the velocity of $\vec{s}_{xy}$.
\begin{equation}
s_{xy} = s \sin \theta
\end{equation}
and
\begin{equation}
\Omega \nu_{0xy} = \Omega s_{0xy} = \frac{|\vec{\mu}||\vec{B}|\sin \theta}{L \sin \theta} = \frac{\mu B}{L}
\end{equation}

We can conclude from this expression that the angular velocity of the rotation
of the particle depends on the magnetic momentum $\vec{\mu}$, the magnetic field
$\vec{B}$ in which the particle is immersed and the spin $\vec{s}$ of the particle,
and it is independent of the relative positions of $\vec{\mu}$, $\vec{B}$ and
$\vec{\nu}_0$.

In the case in question, the angular velocity of the rotation of the particle
in $\vec{B}$ is expressed by:
\begin{equation}
\vec{\Omega} = \gamma \frac{e}{2m} \vec{B}
\end{equation}

According to the theory developed above, the particle will draw a circular
trajectory. We can determine the radius of the orbit drawn by the electron
under the influence of a magnetic field
\begin{equation}
r = \frac{\nu_0}{\Omega} = \frac{2 m \nu_0}{\gamma e B}
\end{equation}

In the specific case in which the particle is an electron, $\gamma$ has a value
of 2 and the radius of the orbit is calculated from the following equation:
\begin{equation}
r = \frac{\nu_0}{\Omega} = \frac{m \nu_0}{e B}
\end{equation}

From the conclusions summarized above, we state that the particle is submitted
to a central force affecting the charged particle inside the magnetic field.
To calculate its value, we only need to recall the formula:
\begin{equation}
\vec{F} = m \vec{\nu}_0 \times \vec{\Omega}
\end{equation}
and substitute $\vec{\Omega}$ for its value previously obtained. Then we get:
\begin{equation}
\vec{F} = m \vec{\nu}_0 \times \gamma \frac{e}{2 m}\vec{B}
\end{equation}

Once again, if we apply this to the case in which the particle is an electron,
the gyromagnetic factor is 2, and hence
\begin{equation}
\vec{F} = e \vec{\nu}_0 \times \vec{B}
\end{equation}

The behaviour predicted by the theory here shown of a charged spinning
particle penetrating into a magnetic field with velocity $\nu$, matches the
one that can be observed in the laboratory, and the force to which it will
be submitted is the well known Lorentz force.

\section*{\normalsize EXPERIMENTS CARRIED OUT}
\subsection*{Airmodel with spinning disc}

The airmodel is provided of a strong angular momentum by mean of a high speed
spinning disc attached to it by exerting different torques on it, with the
built in flight systems, we achieve modifications in its flight trajectory.

The remote control flying model was designed specifically to verify this
theory (See Fig. 3).
\vspace{10cm}
\begin{center}
\includegraphics[width=7cm, height=4cm]{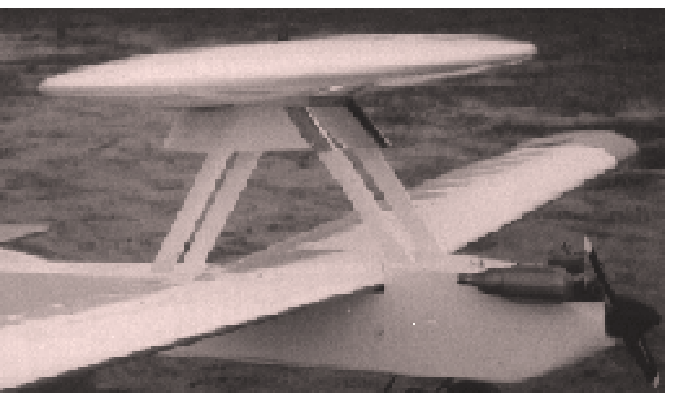}

\vspace{0.5cm}

\includegraphics[width=7cm, height=4cm]{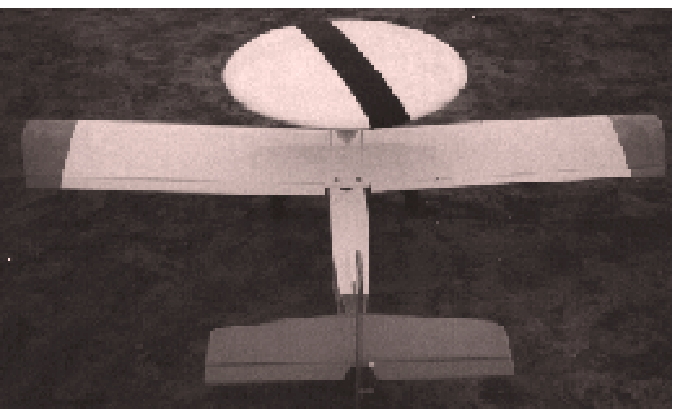}
\end{center}
\begin{center}
\footnotesize{Figure 3. Airmodel with spinning disc}
\end{center}

Basically it consists of a disk that can rotate, attached on top of the
flying model, which provides the necessary support and speed.

Manoeuvring the flying system, aileron to turn right and left, elevators to
shift upwards and downwards and rudder, we can obtain the adequate torques
to affect this motion.

The flying model is controlled by means of a radio equipment affecting the
normal moving elements and the rotation of the disk.

The fist flight was performed without the disk to confirm the normal evolution
of the model.

Once the normal behaviour of the model in this conditions was confirmed, the
disk was attached on top of the model.

With the disk still the system behaves in a similar way that it did without
the disk, although the stability was slightly affected.

With the disk rotation pointer clock way when we actuate the ailerons to the
left, instead of drawing a horizontal circumference, it actually goes up even
to the point of performing a loop.

Actuating the elevators upwards causes the model to turn right, when we actuate
the elevators downwards the model turns to the left.

\subsection*{Spinning top with double suspension assembly}

In this experiment it is proved that the spinning top cam precess with a non zero
radius.

It can be observed as well its extraordinary sensitivity to the torques applied
on it, when these torques have the same direction of the vertical axis, by
modifying the angle between the symmetry axis of the spinning disc and the earths
plane.

For this experiment, a top is mounted with double suspension (See Fig. 4) whereby
it can behave according to the general theory or adopt the type of motion predicted
by the classical treatment.
\begin{center}
\includegraphics[width=7cm, height=6cm]{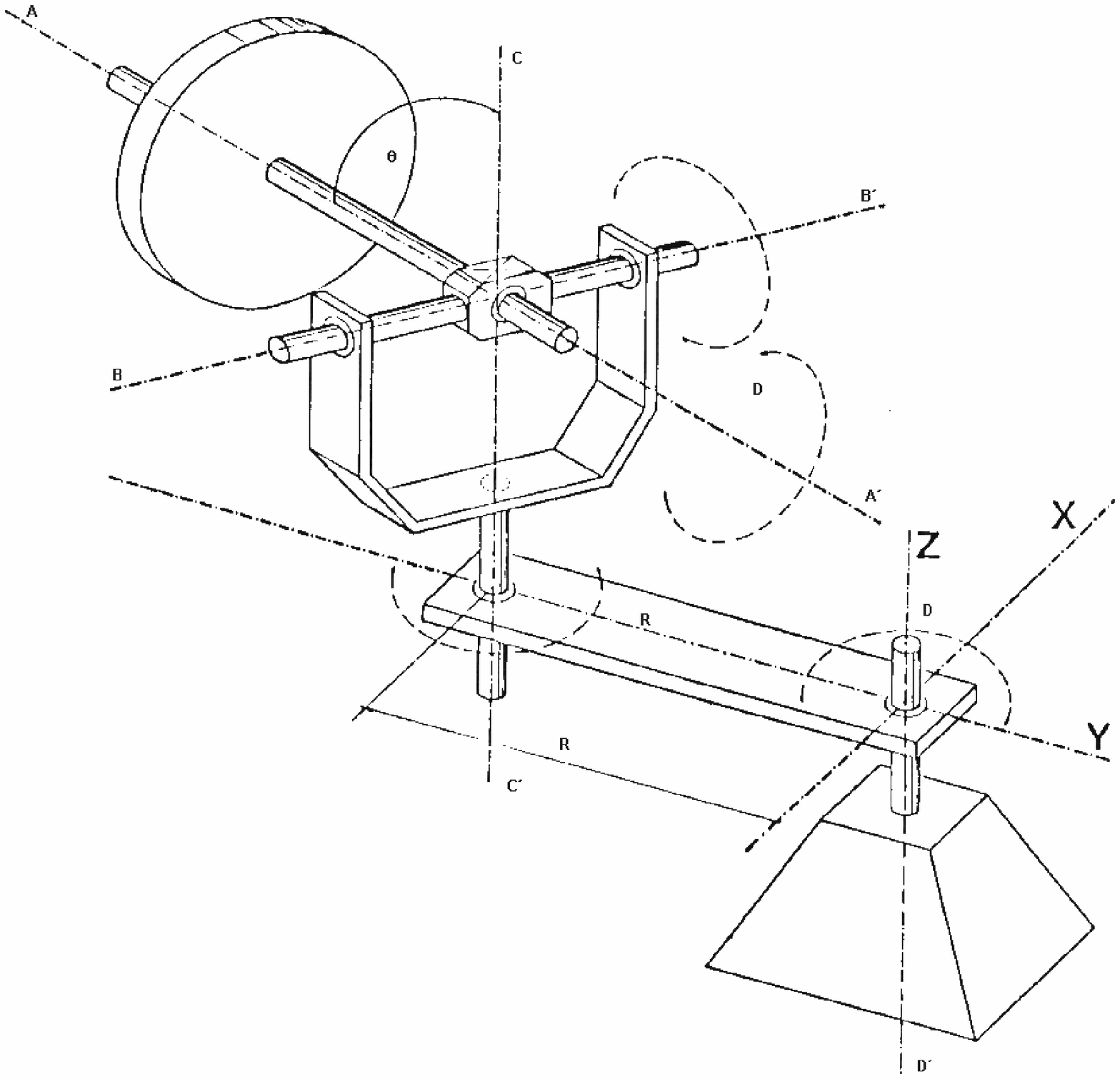}
\end{center}
\begin{center}
\footnotesize{Figure 4. Top mounted with double suspension.}
\end{center}

The top is allowed four possible degrees of rotational freedom:
\begin{enumerate}
\item[a)] $AA'$ axis: The top rotates about its axis of symmetry.
\item[b)] $BB'$ axis: The top will rotate about this axis if nutation is
caused.
\item[c)] $CC'$ axis: The top will rotate about this axis if there is precession.
\item[d)] $DD'$ axis: The system will rotate about this axis if it behaves
according to this theory when we fixed the $CC'$ axis to the support.
\end{enumerate}

The gravitational torque follows the direction of the $BB'$ axis and it produces
an increase in angular momentum $\Delta \vec{L}$ which is vectorially added to
the angular momentum $\vec{L}$ of the top, which is in the direction of the
$AA'$ axis.

Logically, the top and both the $AA'$ and $BB'$ axes could occupy any position
with respect to the support frame.

We provide the top with a constant slow velocity, $\vec{\nu}_0$. Due the top being
mounted on the support frame, the system will rotate about the $DD'$ axis. As
$\vec{\nu}_0$ is slow, the centripetal acceleration due to rotation can be
considered negligible.

According to the classical explanation, the top should precess and even achieve
a nutation motion on the support, with an independent slow rotation motion of the
system about the $DD'$ axis.

The general theory establishes that the velocity of the top, $\vec{\nu}_0$, the
orbital radius and the precession velocity of the top are related by the following
expression:
\begin{displaymath}
R = \frac{\nu}{\Omega}.
\end{displaymath}

Surprisingly this means that if $\Omega$ and $R$ are constant (the support
frame is rigid) there would be only one velocity $\vec{\nu}_0$ for the
movement of the top and a single rotation velocity for the top-support system
about the $DD'$ axis, independently of the impulse given to the support in an
attempt to achieve the desired velocity.

For our experiment, we create a rotation velocity for the top about its axis
of symmetry, this axis ($AA'$) being maintained initially in a horizontal
position although it can really be in any position.

When the top has reached a high rotation velocity about its axis of symmetry,
the support is given an impulse so that the top reaches a velocity $\vec{\nu}$
and the system is then left to evolve freely. For now on, it is also under the
influence of a gravitational torque.

If the system behaves according to classical theory, precession of the top should
not occur with respect to the support.

If on the other hand its behaviour is that described by this theory, precession 
should occur with respect to the $DD'$ axis.

The results of the experiment confirm that the top, in agreement with the theory,
move about the $DD'$ axis with a velocity
\begin{displaymath}
\nu_0 =R \times \Omega_0
\end{displaymath}
and is independent of the impulse it may have received.

In fact the following situations can occur:
\begin{enumerate}
\item[a)] That the impulse received may be bigger than that required to reach
a velocity $\vec{\nu}_0$. The velocity of movement of the top will in all cases
be $\vec{\nu_0}$. However, the angle formed by the axis of symmetry ($AA'$) and the
vertical axis ($CC'$) decreases and there is an increase in potential energy
\end{enumerate}

It should be remembered that $\vec{\Omega}$ is constant and independent of the
angle formed by the $AA'$ and $CC'$ axes while $\vec{L}$ remains constant.
\begin{enumerate}
\item[b)] That the impulse received is equal to that required to reach a velocity
$\vec{\nu}_0$. In this case the position of the top does not vary with respect
to the support. The system rotates about the $DD'$ axis.

\item[c)] That the impulse received is less than that required to reach a velocity
$\vec{\nu}_0$. In this case the system continues to move with a velocity 
$\vec{\nu}_0$, but the angle formed by the $AA'$ and the $CC'$ axes increases.
\end{enumerate}

\subsection*{Pendulum with internal angular momentum}

Two experiments are performed with this device:
\begin{enumerate}
\item[a)] Measure the curvature of the trajectory at the very first instants
of its evolution, where we assess that it is close to a circular trajectory.

\item[b)] Check that the final trajectory is an elliptical one in which the
perihelio of the orbit advances.
\end{enumerate}

This experiment is performed with a pendulum that consist of a fibre-glass 
sphere. The sphere houses a disk that can rotate around its transversal axis
(See Fig. 5).

The pendulum is suspended from a wire attached to a point one centimeter distant
of the edge of the disk shaft.

If we observe the evolution of the pendulum when the inner disk stands still,
it is the ordinary one.

We force the disk to rotate until the maximum speed is reach. At that moment we
let the sphere free. The system is under the influence of a torque caused by its 
weight and the wire strain. If we measure the trajectory at the very first
instants of its evolution, it happens to be a round one.

Letting the sphere to move, after while it follows an elliptical orbit, on which
the perihelium shifts in time, in good agreement with this theory (See Fig. 6).

This phenomenon is similar to the one occurring in the shift of a planet orbit
perihelium. And it is similar, as well, to the Larmor precession of a charge in
the presence of a magnetic field.
\begin{center}
\includegraphics[width=3.5cm, height=8cm]{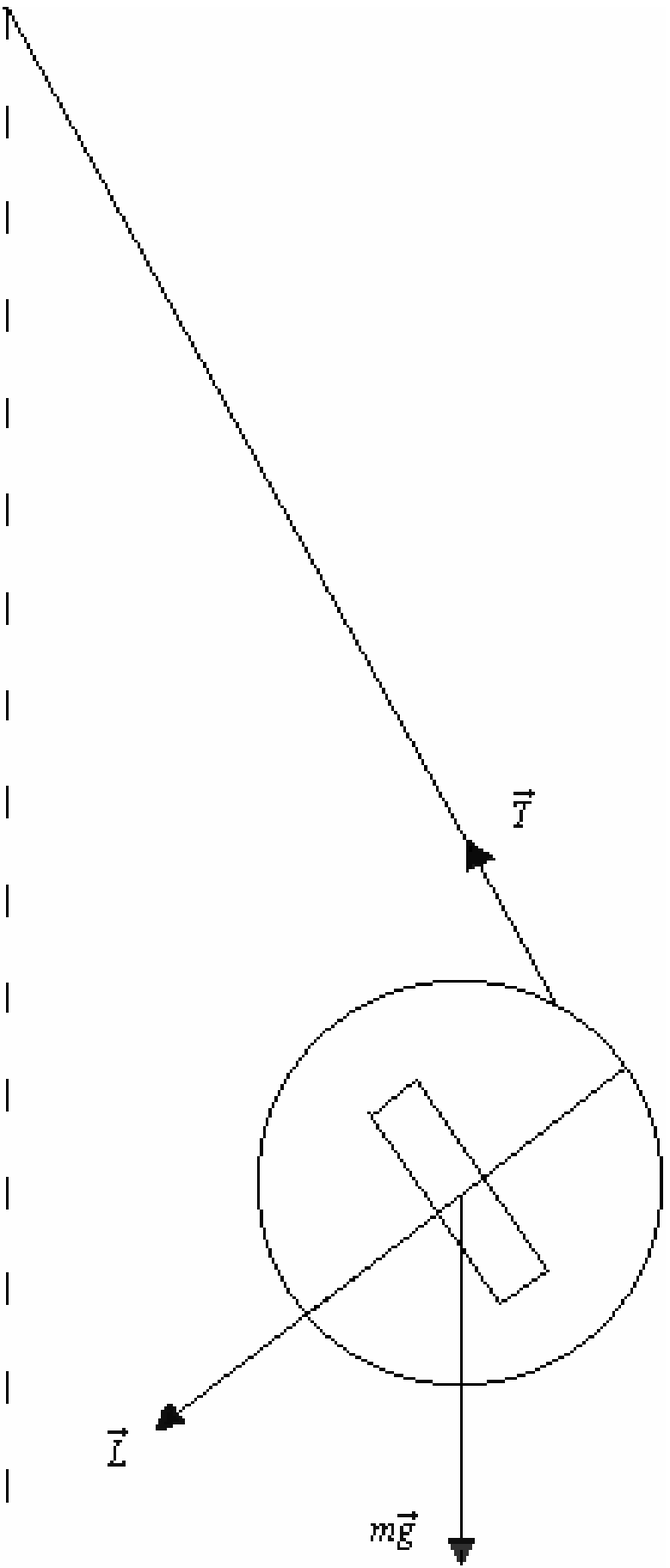}
\end{center}
\begin{center}
\footnotesize{Figure 5. Pendulum with internal angular momentum.}
\end{center}

As in the previous experiment, if we reverse the direction of the disk rotation
the orbit perihelium will also shift in the opposite way that it did before.

\begin{center}
\includegraphics[width=6cm, height=4cm]{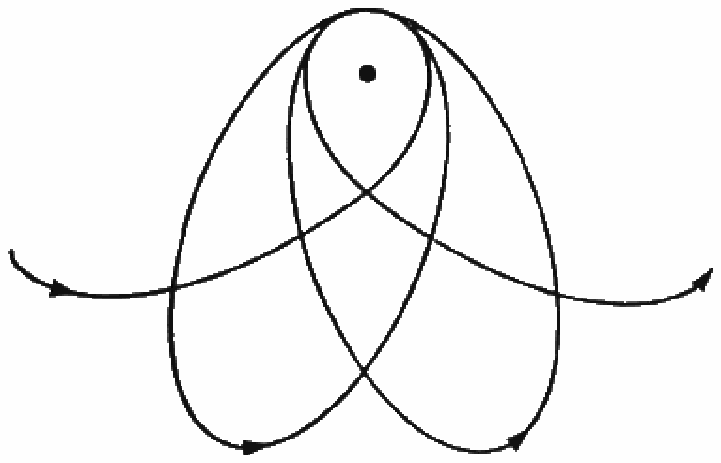}
\end{center}
\begin{center}
\footnotesize{Figure 6. Perihelium shift}
\end{center}

\emph{Theoretical explanation}

\begin{eqnarray}
\frac{d \vec{L}}{dt} &=& \vec{M} \\
\vec{M} &=& \vec{r} \times \vec{F}
\end{eqnarray}

In agreement with this theory, the pendulum is affected by the force
\begin{equation}
\vec{F} = m \vec{\nu} \times \vec{\Omega}
\end{equation}

The torque will be
\begin{equation}
\vec{M} = \vec{r} \times m \vec{\nu} \times \vec{\Omega} = \vec{L} \times \vec{\Omega}
\end{equation}
(It can be proved that the associative property of the vectorial product is
satisfied in this case)

Substituting (45) in (42) we obtain:
\begin{equation}
\frac{d\vec{L}}{dt} = \vec{L} \times \vec{\Omega}
\end{equation}

This one is the equation of the motion of a vector with constant module $|\vec{L}|$,
that precesses around the axis defined by $\vec{\Omega}$, with angular velocity
$|\vec{\Omega}|$.

\vspace{0.5cm}
\noindent
\emph{Larmor Effect} (Samples under the influence of magnetic fields)

In agreement with this theory, an electron embedded inside an atom in the
presence of a magnetic field $\vec{B}$, is submitted to the central force
\begin{equation}
\vec{F} = m \vec{\nu} \times \vec{\Omega}
\end{equation}
which adopts the expression
\begin{equation}
\vec{F} = m \vec{\nu} \times \vec{\Omega} = m \vec{\nu} \times \frac{q \vec{B}}{m c} = \frac{q}{c} \vec{\nu} \times \vec{B}
\end{equation}
and is known as the Lorentz Force.

As the most general expression, we can write:
\begin{equation}
\frac{d \vec{L}}{dt} = \vec{r} \times m \vec{\nu} \times \vec{\Omega} = \vec{L} \times \vec{\Omega}
\end{equation}
(It can be proved that the associative property of the vectorial product is
satisfied in this case)

This is the equation for the motion of $\vec{L}$, which rotates around the vector
$\vec{B}$ with an angular velocity:
\begin{equation}
\vec{\Omega} = -\frac{q}{mc} \vec{B}
\end{equation}

The general expression of the central force, which includes the Lorentz force 
as a particular case, allows us to explain the Larmor effect and its meaning in
an easy and accurate way within the range of a wider and more important 
phenomenon.

\subsection*{Magnetic spinning top}

With this experiment it is proved that the direction way of the precession can
be manipulated in accordance with the torque applied.

For the this experiment we have a magnetized rod, a aluminium cone and a magnet.

We attach the rod to the cone. We have build up a magnetized spinning top
(See Fig. 7).
\begin{center}
\includegraphics[width=7cm, height=5cm]{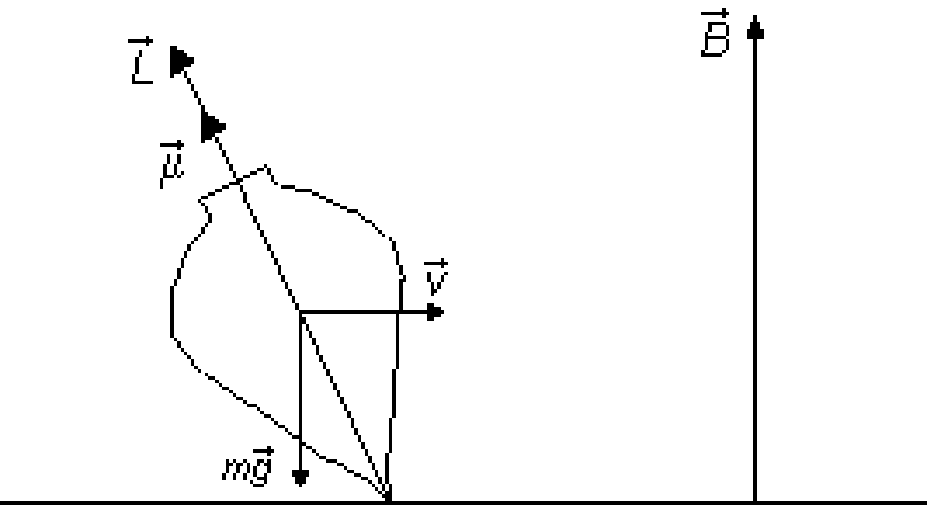}
\end{center}
\begin{center}
\footnotesize{Figure 7. Magnetic spinning top.}
\end{center}

We observe the behaviour of the magnetized spinning top in the presence of a
magnetic field. The interaction between both fields produces a torque which
affects the spinning top making him to draw a circular trajectory.

The spinning top will not collapse on the magneto as long as it has an
angular momentum.

\section*{\normalsize EXPERIMENT SUGGESTED TO TEST THIS THEORY: INTERACTION OF A HOMOGENEOUS MAGNETIC FIELD AND A PERPENDICULAR GYRATING MAGNETIC FIELD, WITH A PARTICLE WITH SPIN AND MAGNETIC MOMENT}

\subsection*{Formulation of the problem}

The particle with spin, $\vec{S}$, moves with velocity $\vec{\nu}$. It first passes
into a homogeneous magnetic field $\vec{B}'$, inducing the magnetic moment,
$\vec{\mu}$, of the particle, polarized in the direction of the field. 
It later enters a region in which a
homogeneous field, $\vec{B}$, and a gyrating field, $\vec{H}_0$, with
rotation frequency, $\dot{\phi}$, are superposed. Fields
$\vec{B}'$ and $\vec{B}$ are parallel and in fact could even be the same field,
although their function is different in each region.

Reference frame (See Fig. 8).

$X Y Z$ Fixed system in the particle but which does not rotate with this.

\hspace{1.5cm} $\vec{B}$ is on the $Z$ axis.

\hspace{1.5cm} $\vec{H}_0$ is on the $X Y$ plane.

$X' Y' Z'$ Fixed frame in the particle which precesses with this.

\hspace{1.5cm} $\vec{S}$ and $\vec{\mu}$ are on the $Z'$ axis.

\hspace{1.5cm} $\vec{\nu}$ is on the $Y'$ axis.

$X_0 Y_0 Z_0$ Reference frame from which observations are made.

\hspace{1.5cm} $\vec{B}$ is on the $Z_0$ axis.

\hspace{1.5cm} $\vec{H}_0$ is on the $X_0 Y_0$ plane.

\begin{center}
\includegraphics[width=7cm, height=6cm]{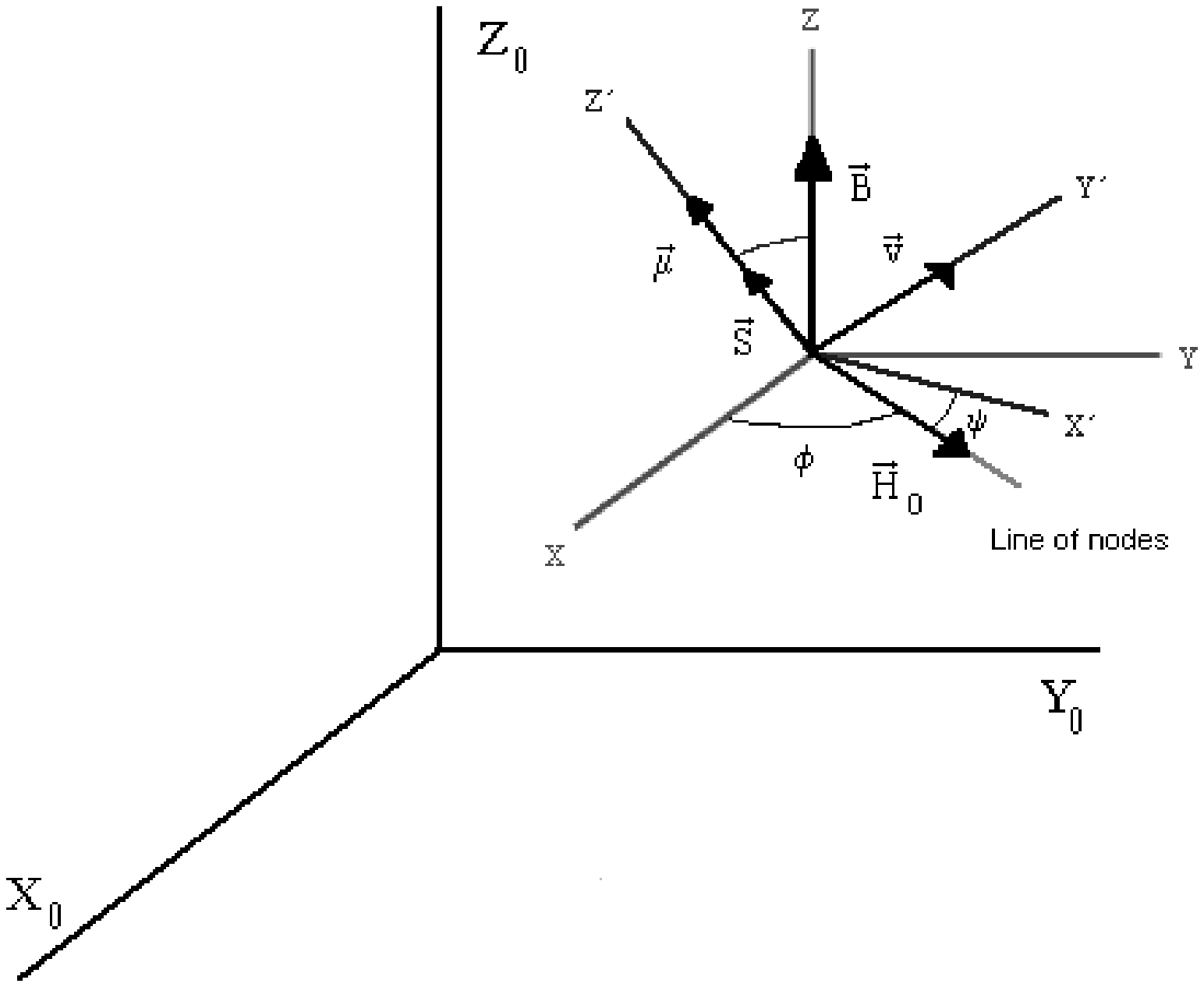}
\end{center}
\begin{center}
\footnotesize{Figure 8. Reference frames.}
\end{center}

The rotation velocity, $\vec{\Omega}$, of the frame of axes linked to the
solid, $X' Y' Z'$, will be expressed within this system as:
\begin{equation}
\vec{\Omega} = \Omega_1 \vec{i}' + \Omega_2 \vec{j}' + \Omega_3 \vec{k}'
\end{equation}

This rotation velocity, $\vec{\Omega}$, can also be expressed with respect
to the frame of reference axes, $X Y Z$ by using Euler angles.
\begin{eqnarray}
\Omega_1 &=& \dot{\phi} \sin \theta \sin \psi + \dot{\theta} \cos \psi \nonumber \\
\Omega_2 &=& \dot{\phi} \sin \theta \sin \psi - \dot{\theta} \cos \psi  \\
\Omega_3 &=& \dot{\phi} \cos \theta + \dot{\psi} \nonumber
\end{eqnarray}

Our particle can also rotate with respect to the frame of axes, $X' Y' Z'$, and
the most general expression to describe this angular velocity is:
\begin{equation}
\vec{\omega} = \omega_{X'} \vec{i}' + \omega_{Y'} \vec{j}' + \omega_{Z'} \vec{k}'
\end{equation}

It is proved that $\dot{\psi} = \omega_{Z'}$. The proof is easy, for if the
$Z'$ axis is fixed, i.e. both $\phi$ and $\theta$ are constant, the rotations
about the $Z'$ axis are rotations of the $X'$ and $Y'$ axes in the $X' Y'$
plane. Hence, the Euler angle that reports the rotation is $\psi$. It can
then be concluded that
\begin{equation}
\dot{\psi} = \omega_{Z'}
\end{equation}

By agreement, but without loss of generality, let us consider that the particle
has an angular momentum, $\vec{S}$, that is constant in module.

\subsection*{Frame of axes linked to the solid}

In the definition of the frame of axes linked to the solid, we can either choose:
\begin{enumerate}
\item[a)] The frame of axes strictly accompanies the particle in the rotation
which, in classical terms, gives its internal angular momentum.

\item[b)] The frame of axes linked to the solid is defined by a parallel axis
to the internal angular momentum of the particle, and the other two are
perpendicular to each other and are in a plane which is perpendicular to the
angular momentum. Internal angular momentum is understood to be a quality of
the particle whose mathematical characteristics coincide with those of the
angular momentum, without considering the physical nature of this quality.
In other words, no hypothesis is formed as to whether or not the angular
momentum implies rotations of the particle under study.
\end{enumerate}

For the purposes of our problem, definition (b) has been used.

However, it should be remembered that our particle can reach a rotation
velocity, $\vec{\omega}$, as a result of the interactions to which it can
be subjected.

\subsection*{Analysis of the interactions}

Given the nature of our problem, the particle penetrates a homogeneous
magnetic field, $\vec{B}$, inducing the magnetic moment, $\vec{\mu}$, of
the particle, polarized in the direction of the field.

Hence, when the particle penetrates the magnetic field, $\vec{B}$ 
(parallel to $\vec{B}'$, and can even coincide with it) and $\vec{H}_0$,
the magnetic moment, $\vec{\mu}$, is only (initially) sensitive to the
gyrating magnetic field, $\vec{H}_0$. This, consequently, will be the
first interaction that we analyze.

\subsection*{Interaction with the magnetic field, $\vec{H}_0$}

The interaction of the gyrating magnetic field, $\vec{H}_0$, with the
magnetic moment, $\vec{\mu}$, does not modify the energy of the system,
as both are permanently perpendicular.

The energy of the interaction is expressed
\begin{equation}
E_H = \vec{H}_0 \cdot \vec{\mu} = H_0 \mu \cos \frac{\pi}{2} = 0
\end{equation}

The interaction between $\vec{H}_0$ and $\vec{\mu}$, is equivalent
to the action of a torque, $\vec{\Gamma}_H$, on the particle, where
\begin{equation}
\vec{\Gamma_H = \vec{\mu} \times \vec{H}_0}
\end{equation}

$\vec{\Gamma}_H$ is perpendicular to the $X'$ and $Z'$ axes, and is
located on the $Y'$ axis.

By using the previously obtained Euler equations (9) and applying
them to our problem:
\begin{eqnarray}
\dot{\omega}_{Z'} &=& \frac{M_3}{I_3} = 0 = \dot{\psi} \nonumber \\
\Omega_1 &=& \frac{\Gamma_H}{S} = \frac{\mu H_0}{S} \\
\Omega_2 &=& \frac{M_1}{S} = 0 \nonumber
\end{eqnarray}
so that $\dot{\psi}$ = 0 in all cases, given that $M_3$ = 0 at all
times. We can avoid $\psi$ by making it equal to zero in all case
(See Fig. 9).

The Euler equations depending on the Euler angles (52), are 
simplified with the result that
\begin{eqnarray}
\Omega_1 &=& \dot{\theta} = \frac{\mu H_0}{S} \nonumber \\
\Omega_2 &=& \dot{\phi} \sin \theta = 0 \\
\Omega_3 &=& \dot{\phi} \cos \theta \nonumber
\end{eqnarray}
from which it is deduced that
\begin{center}
\includegraphics[width=7cm, height=6cm]{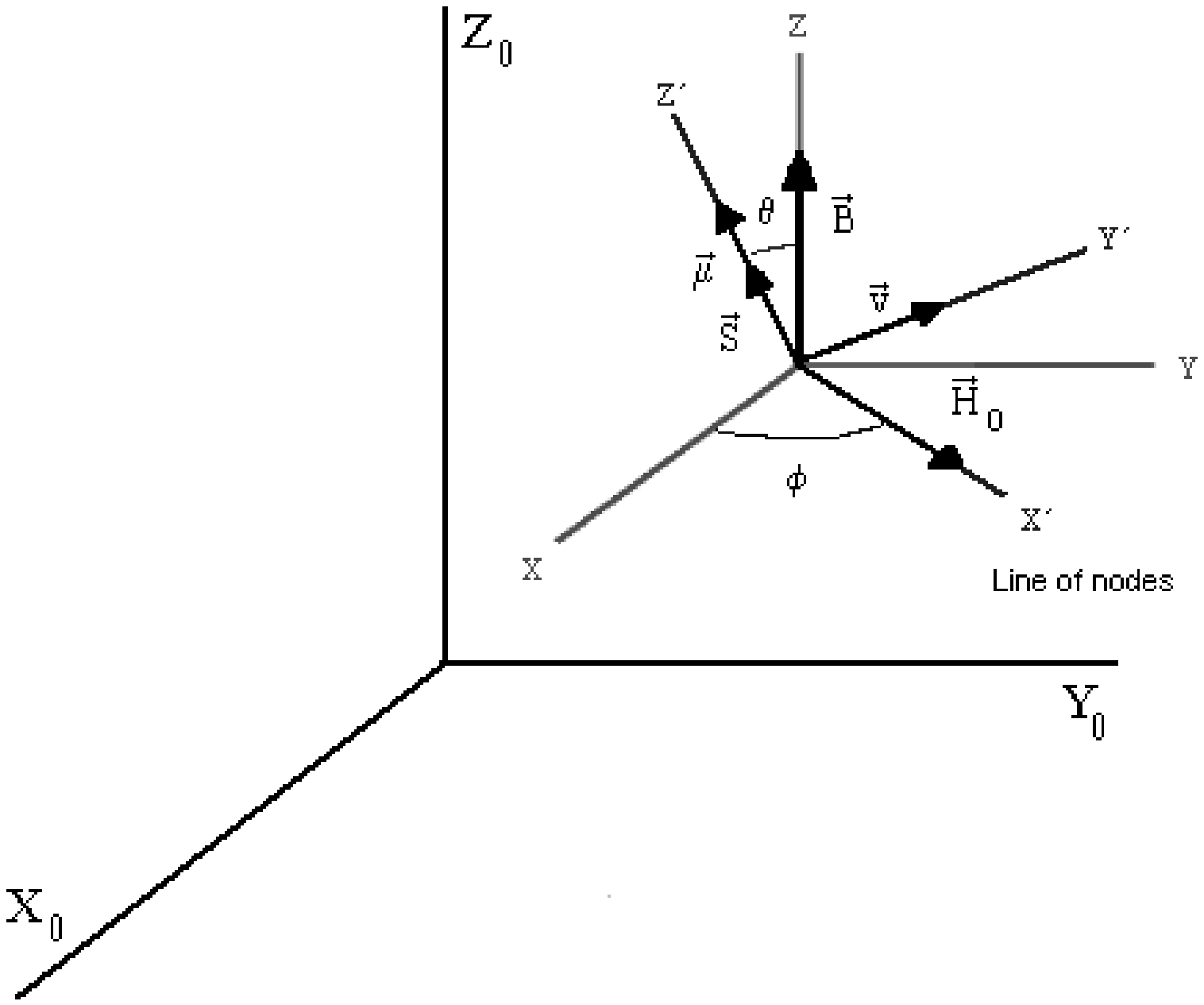}
\end{center}
\begin{center}
\footnotesize{Figure 9. Interaction with the magnetic field, $\vec{H}_0$.}
\end{center}

\begin{eqnarray}
\vec{\theta} &=& \frac{\mu H_0}{S} \nonumber \\
\dot{\phi} &=& 0 \\
\Omega_3 &=& 0 \nonumber
\end{eqnarray}
in other words, the rotation of the frame linked to the solid on
$X' Y' Z'$, $\Omega_1$, is constant and equal to $\dot{\theta}$, where
$\dot{\theta}$ is the rotation of the frame of axes linked to the solid
with regard to $X Y Z$ expressed in the Euler angles.

We have obtained the rotation velocity, $\dot{\theta}$, but we have to
calculate the rotation radio $r(t)$ if we want to know the trajectory.
To do this we will use the following equality between differential
operators,
\begin{displaymath}
\frac{d}{dt} = \frac{d^*}{dt} + \vec{\Omega} \times \hspace{0.7cm},
\end{displaymath}
where \\
$\frac{d}{dt}$ is the derivative in the inertial frame, in our case
$X_0 Y_0 Z_0$, \\
$\frac{d^*}{dt}$ is the derivative in the non-inertial frame, in our
case $X' Y' Z'$, \\
$\vec{\Omega}$ is the angular rotation velocity of the frame of axes
linked to the solid in relation to the inertial frame of axes.

We will apply this equality to $\vec{r}$, a vector of position of one
frame with respect to another, and for convenience we will express
all the vectors in the $X' Y' Z'$ frame.
\begin{equation}
\frac{d^* \vec{r}}{dt} = \frac{d \vec{r}}{dt} - \vec{\Omega} \times \vec{r}
\end{equation}

If we represent the equation in components, we have
\begin{eqnarray}
\dot{r}_{X'} + r_{Z'} \Omega_2 - r_{Y'} \Omega_3 &=& \nu_{X'} \nonumber \\
\dot{r}_{Y'} + r_{X'} \Omega_3 - r_{Z'} \Omega_1 &=& \nu_{Y'} \\
\dot{r}_{Z'} + r_{Y'} \Omega_1 - r_{X'} \Omega_2 &=& \nu_{Z'} \nonumber
\end{eqnarray}

Taking into account the initial conditions (See Fig. 10), $\Omega_2$ = 0
and $\Omega_3$ = 0.
\begin{center}
\includegraphics[width=7cm, height=6cm]{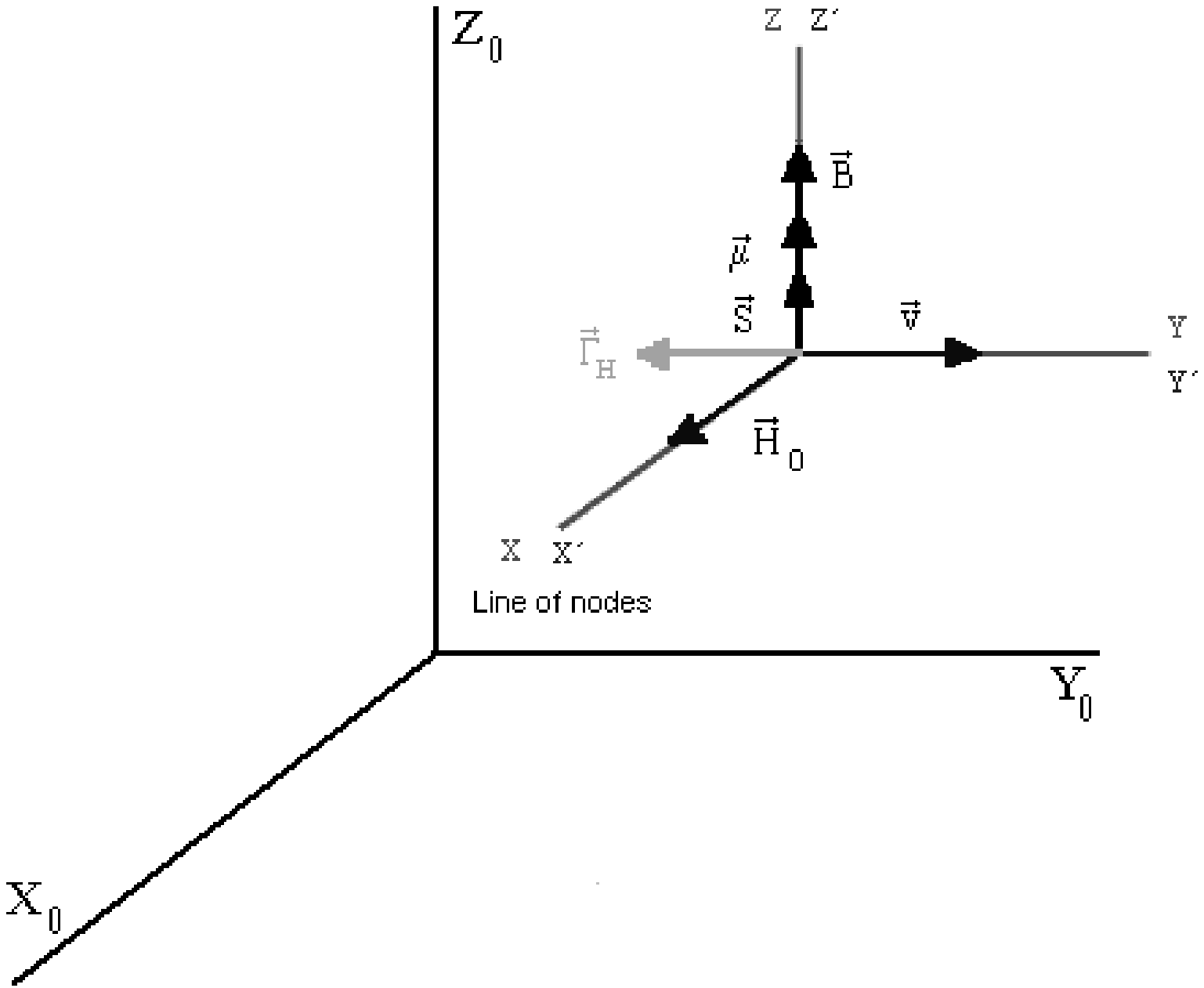}
\end{center}
\begin{center}
\footnotesize{Figure 10. Initial conditions.}
\end{center}

The interaction does not modify the energy of the system, as explained
beforehand, so the module of $\vec{\nu}$ has to remain constant.
\begin{displaymath}
|\vec{\nu}| = \nu = \mbox{constant.}
\end{displaymath}
and, $\nu_{X'}$ = 0, $\nu_{Y'}$ = $\nu$, $\nu_{Z'}$ = 0.

As the energy of the system cannot vary, 
$\dot{r}_{X'}$ = $\dot{r}_{Y'}$ = $\dot{r}_{Z'}$ = 0. If they differed
from zero tangential accelerations would be involved and therefore a 
variation in the total energy of the system.

These conditions simplify the equations of our system (61) and we
obtain
\begin{equation}
r_{Z'} = -\frac{\nu}{\Omega_1}
\end{equation}

This is to say, the particle and the frame of axes linked to the particle
describe a circular trajectory with respect to the observation frame. The
movement plane is perpendicular to $\vec{H}_0$.

\subsection*{Interaction with the magnetic field $\vec{B}$}

At the moment the circular trajectory is commenced, $\theta$ is no longer
null and the interaction with the field, $\vec{B}$, begins (See Fig. 11).
\begin{eqnarray}
\vec{\Gamma} &=& \vec{\mu} \times \vec{B} \\
\Omega_2 &=& \frac{\mu B \sin \theta}{S}
\end{eqnarray}
\begin{center}
\includegraphics[width=7cm, height=6cm]{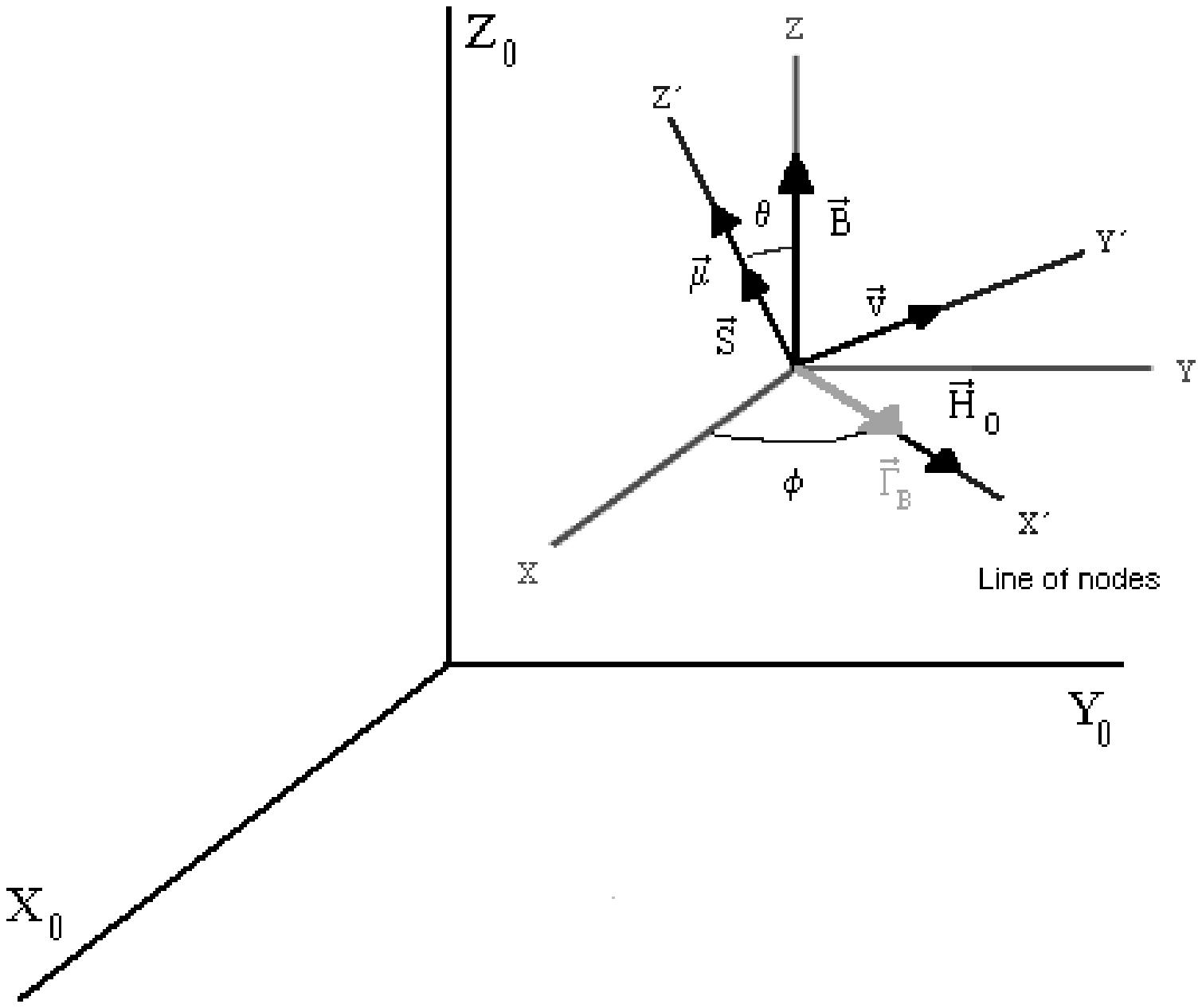}
\end{center}
\begin{center}
\footnotesize{Figure 11. Interaction with the magnetic field.}
\end{center}

$M_3$ continues to be null and, therefor, so do $\dot{\psi}$ and
$\psi$, which do not vary, but $\Omega_2$ is now different from
zero.
\begin{eqnarray}
\Omega_1 &=& \dot{\theta} \nonumber \\
\Omega_2 &=& \dot{\phi} \sin \theta = \frac{\mu B \sin \theta}{S} \\
\Omega_3 &=& \dot{\phi} \cos \theta \nonumber
\end{eqnarray}

We finally obtain,
\begin{eqnarray}
\dot{\theta} &=& \frac{\mu H_0}{S} \nonumber \\
\dot{\phi} &=& \frac{\mu B}{S} \\
\Omega_3 &=& \frac{\mu B \cos \theta}{S} \nonumber
\end{eqnarray}

These are the expressions of the rotation of the frame of axes linked
to the solid in the base that defines the trihedral linked to the solid.

We can also express this same rotation velocity of the frame of axes
linked to the solid with respect to the reference frame, by means of
the Euler angles, but in the base defining the trihedral of
reference frame.
\begin{eqnarray}
\omega_X &=& \dot{\theta} \cos \phi + \dot{\psi} \sin \theta \sin \phi \nonumber \\
\omega_Y &=& \dot{\theta} \sin \phi - \dot{\psi} \sin \theta \cos \phi  \\
\omega_Z &=& \psi \cos \theta + \dot{\phi}  \nonumber
\end{eqnarray}

As $\dot{\psi}$ = $\psi$ = 0, we obtain
\begin{eqnarray}
\omega_X &=& \dot{\theta} \cos \phi \nonumber \\
\omega_Y &=& \dot{\theta} \sin \phi \\
\omega_Z &=& \dot{\phi} \nonumber
\end{eqnarray}

These expressions clearly reflect the behaviour of the frame of axes
linked to the solid and of the solid, with respect to the reference
axes.

Visualization of the movement is assisted by considering the following
specific situations $\dot{\phi}$ = 0
\begin{enumerate}
\item[i)] For $\phi$ = 0
\begin{eqnarray*}
\omega_X &=& \dot{\theta} \\
\omega_Y &=& 0 \\
\omega_Z &=& 0 \\
\end{eqnarray*}
\end{enumerate}
We would have a rotation about the $X$ axis.
\begin{enumerate}
\item[j)] For $\phi$ = $\frac{\pi}{2}$
\begin{eqnarray*}
\omega_X &=& 0 \\
\omega_Y &=& \dot{\theta} \\
\omega_Z &=& 0 \\
\end{eqnarray*}
\end{enumerate}
We would have a rotation about the $Y$ axis.

The rotation velocity in the $XY$ plane is
\begin{eqnarray}
\vec{\omega}_{XY} &=& \dot{\theta} (\cos \theta \vec{i}+ \sin \theta \vec{j}) \nonumber \\
\\
|\vec{\omega}_{XY}| &=& \dot{\theta} \nonumber
\end{eqnarray}

Therefore, the total movement is a circular trajectory of radius, $r_{Z'}$,
perpendicular to the field, $\vec{H}_0$, which in turn spins about the
field, $\vec{B}$, with velocity $\dot{\phi}$ (classically know as Larmor
rotation frequency). This explains why the frequency of the oscillating
field, $\vec{H}_0$, has to coincide with the Larmor frequency.

The behaviour of the particle indicates that it is subjected to two
central forces
\begin{equation}
\vec{F}_1 = m \vec{\nu} \times \vec{\Omega}_1 = m \vec{\nu} \times \dot{\theta} \vec{i}'
\end{equation}                                                                           
\begin{equation}
\vec{F}_2 = m \vec{\nu} \times \dot{\phi} \vec{k}
\end{equation}                                                                           

From an energy point of view, we can confirm that the magnetic field,
$\vec{H}_0$, does not modify the energy of the system, and that the
magnetic field, $\vec{B}$, cyclically modifies the potential energy of
the particle, since
\begin{eqnarray}
E_B &=& \vec{B} \cdot \vec{\mu} \cos{\theta} \nonumber \\
\\
\mbox{with}& &\theta(t) = \frac{\mu H_0}{S}t \nonumber
\end{eqnarray}
but with its kinetic energy unvaried.

\subsection*{Additional considerations}

The reader will have already noted the similarity between this behaviour
and the Larmor effect.

It is know that this effect is produced by applying a magnetic field,
$\vec{B}$, to a particle of charge, $q$, moving in an orbit around a
fixed specific charge, $q'$. The result is a precession of the trajectory
around the direction of the applied magnetic field, with a precession
velocity, $\omega_L$ = $\frac{qB}{2m}$, know as Larmor frequency, with
the proviso that the cyclotronic frequency is directly obtained by
this procedure.

This statement can be expressed by saying that the angular momentum
vector of the particle with respect to the rotation axis, referred to
as orbital angular momentum, precesses about the direction of the
magnetic field, $\vec{B}$.

From our procedure, it is particularly easy to reach the same
conclusions by considering that
\begin{eqnarray}
\frac{d \vec{L}}{d t} &=& \vec{\Gamma}_B = \vec{r} \times \vec{F} \nonumber \\
\\
\vec{L} &=& \vec{r} \times m \vec{\nu} \nonumber
\end{eqnarray}

By substituting our development, $\vec{F} = m \vec{\nu} \times \vec{\Omega}$,
in (73) we obtain
\begin{eqnarray}
& &\frac{d \vec{L}}{d t} = \vec{r} \times (m \vec{\nu} \times \vec{\Omega}) \nonumber \\
\\
&=&(\vec{r} \times m \vec{\nu}) \times \vec{\Omega} = \vec{L} \times \vec{\Omega} \nonumber
\end{eqnarray}

\twocolumn[
\begin{center}
{\includegraphics[width=16cm,height=6.5cm]{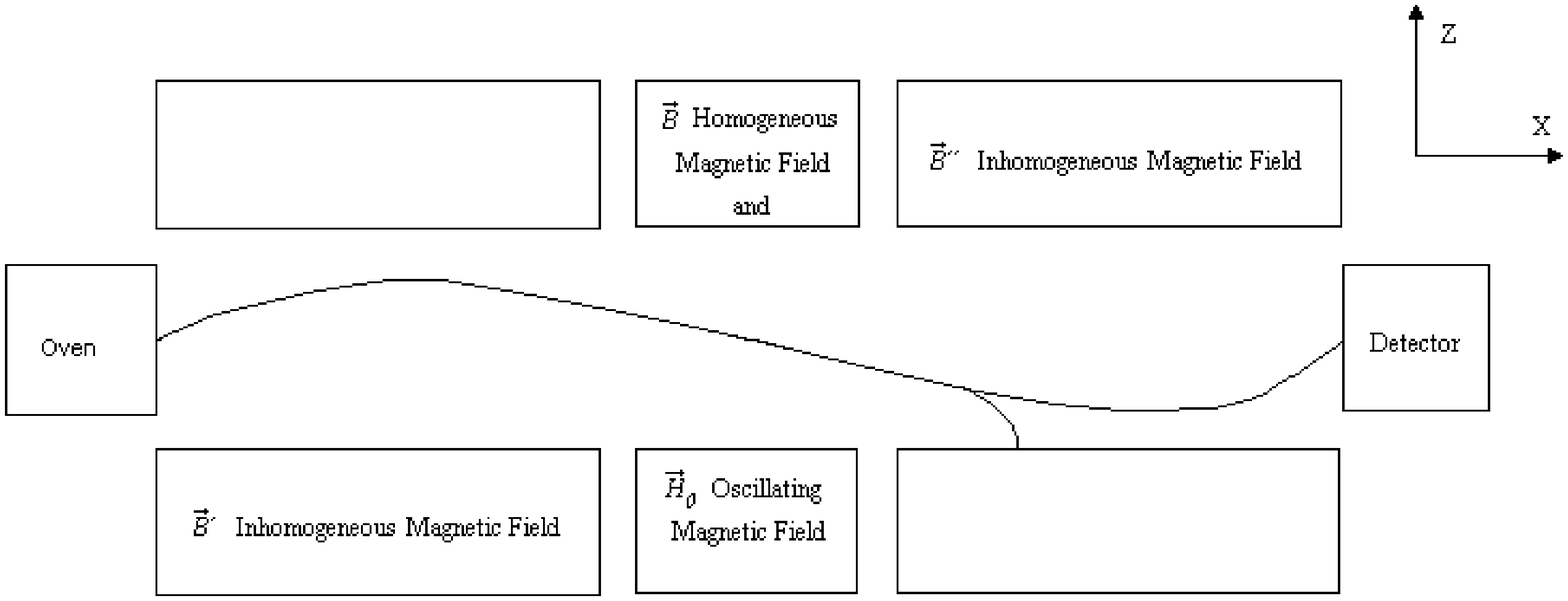}}
\\
\footnotesize{Figure 12. Rabi's experiment.}
\\
{\includegraphics[width=16cm,height=6.5cm]{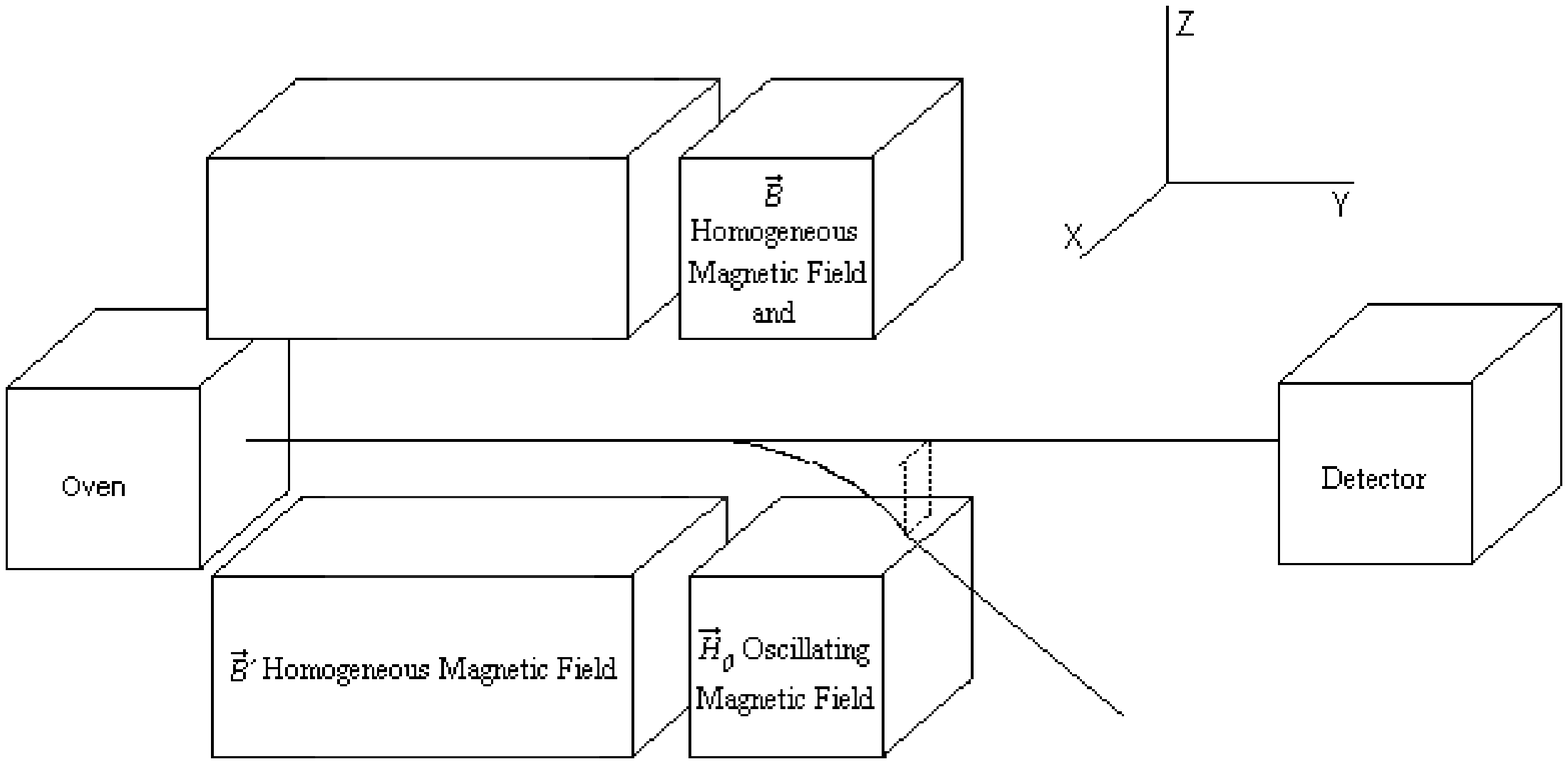}}
\\
\footnotesize{Figure 13. Experiment suggested.}
\end{center}
]

(the triple vector product does not generally fulfill the associative property,
but it can be demostrated in our.)

Equation (74) describes the angular momentum precession about the magnetic
field with a precession rate $\vec{\Omega}$.

The Larmor effect results from subjecting the particle to the Lorentz
central force, because it is within $\vec{B}$. From the following, the
Lorentz force is equivalent to our force, given that
\begin{equation}
F = m \nu \Omega = m \nu \frac{\mu B}{S} = m \nu \frac{e}{m}B = e \nu B
\end{equation}

Which demostrates that both movements are equivalent.

\subsection*{Experimental conditions}

This experiment can be carried out by emulating the rotating magnetic
field using a radio frequency magnetic field, just as Rabi did in his
experiments.

Rabi's experiment (see References (2)) consists of a collimated
particle beam that crosses an inhomogeneous magnetic field. It later passes
through a region where a homogeneous and a radio frequency magnetic field
are superposed, and finally passes through an inhomogeneous field that
refocuses the beam towards the detector.

The inhomogeneous fields separate the beam into different beams according
to their magnetic moment (the dependence this on the spin) as in an

experiment of Stern-Gerlach. When leaving the first field, these beams
later pass into the second inhomogeneous field, which refocuses the
beams to the detector. By adjusting the second inhomogeneous magnetic
field we will obtain refocusing conditions for a beam or group of
beams.

In the central part of the experimental arrangement, the homogeneous
magnetic field is superposed on the radio frequency field. In this
region, the spin and magnetic moment are deflected with respect to the
constant field when the oscillating field frequency approaches the
Larmor precession frequency. This process is know as nuclear magnetic
resonance.

After deflection, the atom is on another level which will not fulfill
the refocusing condition and a decrease in the beam intensity will be
observed in the detector. This procedure is used to study nuclear spin,
nuclear magnetic moments and hyperfine structures.
However, the behaviour of the particles in Rabi's experiment is theoretically
justified in our exposition and it is therefore unnecessary to use
inhomogeneous magnetic fields (as is required in Rabi's experiments).

Our experiment only requires the presence of a homogeneous magnetic field
to induce the magnetic moment, $\vec{\mu}$, of the particle, polarized
in the direction of the field.

In such a way that:
\begin{enumerate}
\item[1$^o$] Only those particles whose Larmor frequency, $\dot{\phi}$,
coincides with the radio frequency will abandon the rectilinear trajectory.

\item[2$^o$] The trajectory described by these particles should coincide
with that of the theoretical solution given in this article while they
remain within the region in which the two fields coexist. According to
this solution, the particle abandons the rectilinear path and the $XY$
plane.

\item[3$^o$] When the particle abandons the region of coexisting fields,
it will follow a rectilinear movement but will not reach the detector.

\item[4$^o$] This procedure facilitates selection of the different angular
moments because of their dependence on the Larmor precession.
\end{enumerate}

\section*{\normalsize CONCLUSIONS}

When a system with internal angular momentum, moving at constant velocity
with respect to a reference frame, is subjected to an interaction of the
type under study, i.e., a torque perpendicular to the internal angular
momentum vector, it will begin to trace a circular path of radius
\begin{displaymath}
r = \frac{\nu}{\Omega}
\end{displaymath}
where $\Omega = \frac{M}{L}$

Its behaviour is equivalent to that produced by subjecting the cylinder
to a central force
\begin{displaymath}
\vec{F} = m \vec{\nu} \times \vec{\Omega}
\end{displaymath}

As it can be seen in figure 14, the trajectory which particle follows
is the trajectory II, while the 

trajectory I is the intuitive one, but
as it has been shown in this paper it is not the real behaviour of the
particle.

\section*{\normalsize ACKNOWLEDGMENTS}

Through the development of this paper I have had the privilege of maintaining
some discussions about its content with Dr. Jos\'e L. S\'anchez G\'omez.

It will also like to thank Miguel Morales Furi\'o for his support and help
to the development of this article.

In the same way, I will like to express my gratitude to Juan Silva Trigo,
Jes\'us Abell\'an, Berto Gonz\'alez Carrera and to all the other people
who has worked in the design and realization of all the experiments describe
in this article, whose enumeration would be too extensive.

I would also like to express that the responsability for any affirmation here
introduced lies exclusively with the author.
\twocolumn[
\begin{center}
{\includegraphics[width=13cm,height=11cm]{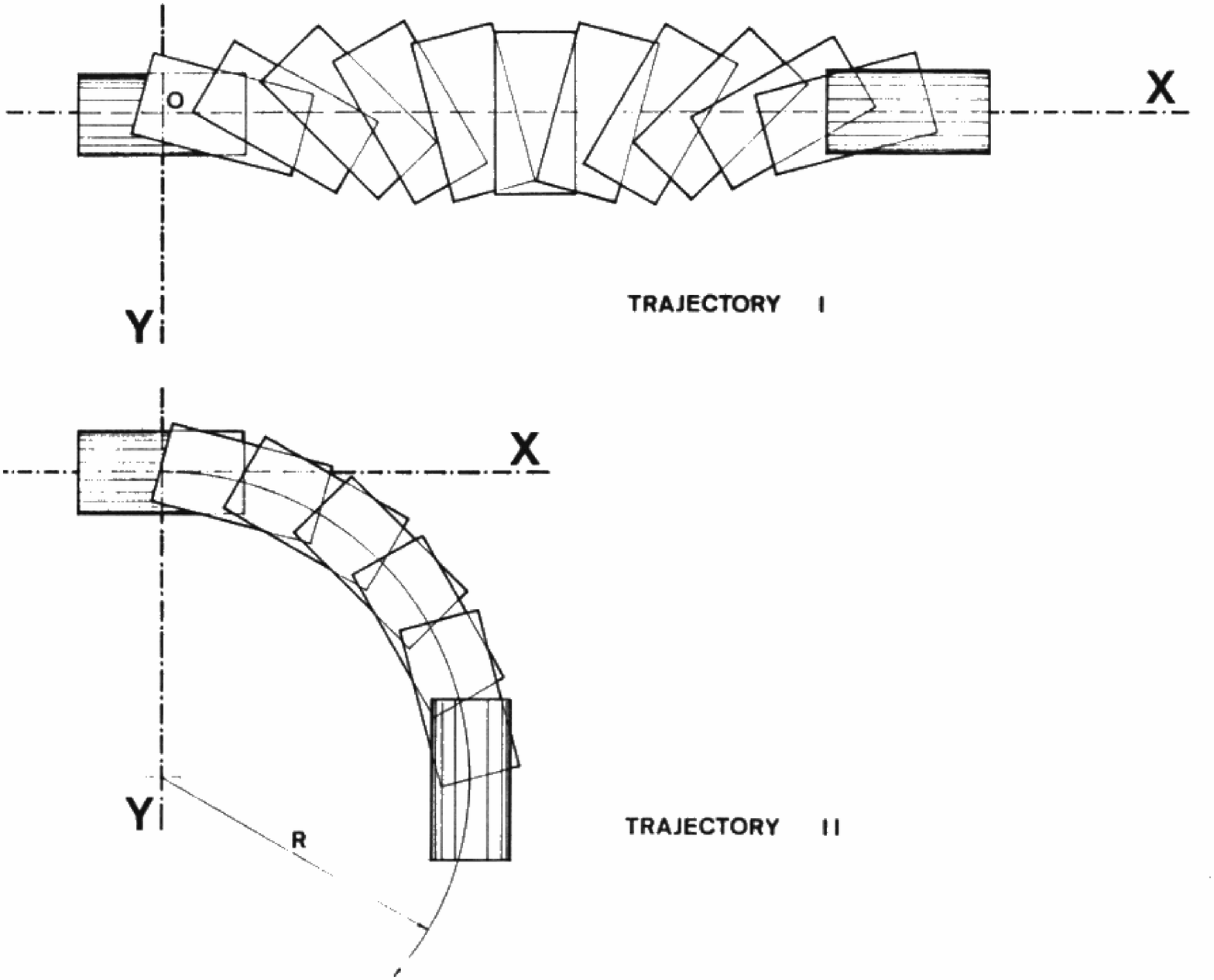}}
\\
\footnotesize{Figure 14. Classical trajectory and trajectory predicted in this article.}
\end{center}
]
\section*{\normalsize REFERENCES}

\begin{enumerate}
\item[[1.a]] FEYNMAN, R. P., LEIGHTON, R. B. AND SANDS, M. \small{ (1964):
\emph{The Feynman Lectures on Physics.} Addison-Wesley.
}

\item[[1.b]] GOLDSTEIN, H. G. \small{ (1970): \emph{Classical Mechanics.}
Addison-Wesley.
}

\item[[1.c]] KIBBLE, T. W. B. \small{ (1966): \emph{Classical Mechanics.}
Mc. Graw-Hill.
}

\item[[1.d]] LANDAU AND LIFSHITZ \small{ (1978): \emph{Mec\'anica.}
Editorial Revert\'e S. A.
}

\item[[1.e]] RAÑADA, A. \small{ (1990): \emph{Din\'amica Cl\'asica.}
Alianza Universidad Textos.
}

\item[[2.a]] KELLOGG, J. M. B., RABI, I. I. AND ZACHARIAS, J. R. 
\small{ (1936) The Gyromagnetic Properties of the Hydrogenes. 
\emph{Physical Review} Vol. {\bf 50}, 472.
}

\item[[2.b]] RABI, I. I., ZACHARIAS, J. R., MILLMAN, S. AND KUSCH, P.
\small{ (1938) A New Method of Measuring Nuclear Magnetic Moment.
\emph{Physical Review} Vol. {\bf 53}, 318.
}

\item[[2.c]] RABI, I. I., ZACHARIAS, J. R., MILLMAN, S. AND KUSCH, P.
\small{ (1939) The Magnetic Moments of $_3Li^6$, $_3Li^7$ and $_9F^{19}$.
\emph{Physical Review} Vol. {\bf 55}, 526.
}

\item[[2.d]] RABI, I. I., ZACHARIAS, J. R., MILLMAN, S. AND KUSCH, P.
\small{ (1938) The Molecular Beam Resonance Method for Measuring Nuclear Magnetic
Moments.The Magnetic Moments of $_3Li^6$, $_3Li^7$ and $_9F^{19}$.
\emph{Physical Review} Vol. {\bf 53}, 495.
}

\item[[2.e]] G\"UTTINGER, P. \small{ (1932) Das Verhalten von Atomen in magnetischen
Drehfeld. \emph{Zeitschrift f\"ur Physik} Bo. {\bf 73}, 169.
}

\item[[2.f]] MAJORANA, E. \small{ (1932) Atomi Orientatti in Campo Magnetico
Variabile. \emph{II Nuovo Cimento} Vol. {\bf 9}, 43.
}

\item[[2.g]] RABI, I. I. \small{ (1936) On the Process of Space Quantization.
\emph{Physical Review} Vol. {\bf 49}, 324.
}

\item[[2.h]] MOTZ, L. AND ROSE, M. E. \small{ (1936) On Space Quantization in
Time Varying Magnetic Fields. \emph{Physical Review} Vol. {\bf 50}, 348.
}

\item[[2.i]] RABI, I. I. \small{ (1937) Space Quantization in
a Gyrating Magnetic Field. \emph{Physical Review} Vol. {\bf 51}, 652.
}

\end{enumerate}

}
\end{document}